\documentclass[aps,prl,twocolumn,showpacs,floatfix]{revtex4}
\usepackage{graphicx}
\usepackage{amssymb}
\usepackage{bm}

\begin{document}

\title{Dynamical Crystallization in the Dipole Blockade of  Ultracold Atoms}
\author{T.\  Pohl$^{1,2}$}
\author{E.\ Demler$^{2,3}$}
\author{M.D.\  Lukin$^{2,3}$}
\affiliation{$^1$Max Planck Institute for the Physics of Complex Systems, N\"othnitzer Strasse 38, 01187 Dresden, Germany}
\affiliation{$^2$ITAMP, Harvard-Smithsonian Center for Astrophysics, 60 Garden Street, Cambridge MA 02138}
\affiliation{$^3$Physics Department, Harvard University, 17 Oxford Street, Cambridge MA 02138}
\date{\today}
\begin{abstract}
We describe a method for controlling many-body states in extended ensembles of Rydberg atoms, forming  crystalline structures during laser excitation of a frozen atomic gas.  Specifically,  we predict the existence of an excitation number staircase in laser excitation of atomic ensembles into Rydberg states.
Each step corresponds to a crystalline state with a well-defined of regularly spaced Rydberg atoms. We show that such states can be selectively excited by chirped laser pulses. Finally, we demonstarte that, sing quantum state transfer from atoms to light, such crystals can be used to create crystalline photonic states and can be probed via  photon correlation measurements. 
\end{abstract}

\maketitle
When cold atoms are excited into high-lying Rydberg states, the resulting interactions give rise to energies that exceed the translational energy by many orders of magnitude and strongly modify the excitation dynamics. In very small ensembles confined to a few micrometers, a single Rydberg atom can entirely block any further excitation.  This, recently observed, "dipole blockade"  \cite{jcz00,lfc01,ujh08,gmw08}, enables the production of highly entangled collective states with potential applications for fast quantum information processing \cite{lfc01,mlw09, mbm08}, and  as single-atom as well as single-photon sources \cite{lfc01} . 

A number of theoretical and experimental studies have addressed the evolution of Rydberg states in large atomic ensembles \cite{tfs04,sra04,crb05,vvz06,rcy08,hrb07,rh05,sc08,wlp08,bmm07,app06}. While strong interactions are critical for many of the predicted and observed phenomena, they also cause a very complex excitation dynamics, having, thus far, prevented a controlled quantum state preparation in large ensembles. For example, experiments \cite{tfs04,sra04,crb05,vvz06,rcy08,hrb07} have probed interesting relaxational excitation dynamics towards equilibrium states whose relevant physics is well captured by mean-field descriptions \cite{tfs04,hrb07,wlp08}.

\begin{figure}[t!]
\begin{center}
\resizebox{0.77\columnwidth}{!}{\includegraphics{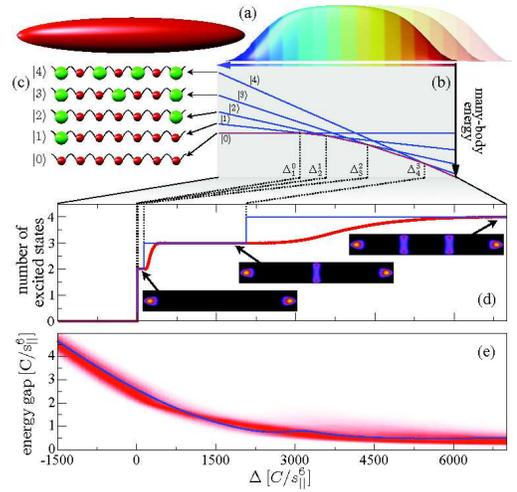}}
\caption{{\bf Adiabatic crystal state preparation.}(a) A cold atomic gas is illuminated by a chirped laser pulse, whose frequency changes from the blue ($\Delta<0$) to the red ($\Delta>0$) side of the driven Rydberg transition. (b) Schematics of the many-body energy spectrum as a function of the laser detuning $\Delta$ and for a Rabi frequency $\Omega=0$. Shown are the lowest-energy states $|n\rangle$ for a given Rydberg atom number $n$. These states correspond to Rydberg atom (green spheres) lattices sketched in (c). Their energy crossings give rise to an excitation number staircase shown in (d) for a chain of $N=31$ atoms (blue curve) and a cigar-shaped condensate (red curve) with the same length $s_{||}$ and $500$ atoms. The insets show the Rydberg atom densities resulting in the latter at the detunings marked by the arrows. Panel (e) shows the ground state energy gap for the chain (blue curve) and the corresponding gap distribution of the condensate (red shading) at $\Omega=10^3\cdot C/s_{||}^6$.\label{fig1}}
\end{center}
\end{figure}

In this work we describe a method that overcomes this obstacle and permits coherent manipulation of strongly correlated, many-body  states of large ensembles. The approach is based on a dynamical crystallization of localized, regularly-spaced collective excitations. The crystalline states can be created with chirped laser pulses, and have a widely tunable lattice spacing. By extending the dipole blockade mechanism \cite{lfc01} to the regime of multiple Rydberg atoms, such excitation crystals provide a suitable initial state to realize quantum random walks \cite{cre06,mba07}. We show that the crystalline correlations can be mapped onto a propagating light beam and be detected via photon correlation measurements as a pulse train of localized single photons. 

The key idea is illustrated in fig.~\ref{fig1}. For repulsive van der Waals interactions and for a given Rydberg atom number $n$, the states with the lowest interaction energy maximize the spacing between Rydberg atoms, which, hence, arrange on regular crystals (fig.~\ref{fig1}b). If initially all atoms are in their ground states, 
for large negative detunings $\Delta$ of the laser field, the initial state coincides with the many-body ground state in the rotating frame of reference. If $\Delta$ is adiabatically increased to positive values, it effectively lowers the energy levels of excited many-body states (fig.~\ref{fig1}c). They ultimately cross at critical detunings $\Delta^{n}_{n+1}$, marking discrete jumps in the Rydberg atom number from $n$ to $n+1$ (fig.~\ref{fig1}d). As a result, larger and larger Rydberg atom crystals are consecutively populated during evolution of the chirped pulse. In essence, the detuning $\Delta$ acts as a control parameter that decreases the energy difference between adjacent number states $|n\rangle$ and $|n+1\rangle$, allowing further Rydberg atoms to "enter" the excitation volume. This interaction blockade is similar to a recently observed effect of interacting atoms in an optical lattice \cite{ctf08}.
Moreover, the resulting "dipole-blockade staircase" bears resemblance to the Coulomb blockade staircase observed in nano-scale solid-state devices \cite{ssd91}, where the number of electrons increases as a function of external lead-voltage. In contrast to such solid state devices, the present system permits dynamical manipulation of strongly interacting particles in a highly coherent manner.

The Rydberg gas is described as an ensemble of two-level systems at fixed positions ${\bf r}_i$, each possessing a ground state $|g_i\rangle$ and an excited state $|e_i\rangle$ \footnote{For the parameters of this work, this essential state picture is well justified for $nS$ Rydberg states, for which near-resoant state mixing \cite{rcy08,yrp09} can be neglected.}, laser-coupled with a Rabi frequency $\Omega(t)$. Within the frozen gas limit, the excitation dynamics is described by the Hamiltonian
\begin{eqnarray} \label{hamiltonian}
\hat{H}=-\Delta\sum_i\hat{\sigma}_{\rm ee}^{(i)}+\sum_{i<j}V_{ij}\hat{\sigma}_{\rm ee}^{(i)}\hat{\sigma}_{\rm ee}^{(j)}+\frac{\Omega}{2}\sum_{i}(\hat{\sigma}_{\rm eg}^{(i)}+\hat{\sigma}_{\rm ge}^{(i)}),\nonumber
\end{eqnarray}
where $\hat{\sigma}^{(i)}_{\alpha\beta}=|\alpha_i\rangle\langle\beta_i|$ ($\alpha_i,\beta_i={\rm e,g}$), and $V_{ij}=C/r_{ij}^6$ describes the repulsive ($C>0$) van der Waals interaction between Rydberg atom pairs at a distance $r_{ij}$. 

We first consider the simplest case of atoms in a one-dimensional optical lattice \cite{vtm05,gme02}, modeled as a chain of length $s_{||}$, that is composed of $N$ atoms equally separated by $a=s_{||}/(N-1)$. 
At zero laser intensity, the general structure of the resulting many-body energy spectrum can be easily understood (see fig.~\ref{fig1}c). The $2^N$ many-body states are composed of $\left({N \atop n}\right)$ states $|n,k\rangle$, where $k$ labels the spatial configuration of $n$ excited Rydberg atoms. Their partially degenerate interaction energies $\epsilon_{n}^{(k)}$ are shifted by $-n\Delta$. Within each $n$-manifold, the state $|n,0\rangle\equiv|n\rangle$ with the lowest energy $E_n^{(0)}\equiv E_n=\epsilon_n^{(0)}-n\Delta$ maximizes the Rydberg atom spacing at $z_{ij}=s_{||}/(n-1)=a_n$, forming a Rydberg atom crystal with lattice spacing $a_n$. 
For small excitation fractions ($a\ll a_n$), the corresponding energies 
\begin{eqnarray}
E_n=-n\Delta+\frac{C}{s_{||}^6}(n-1)^6\sum_j^n\frac{n-j}{j^6}\approx -n\Delta+\frac{C}{s_{||}^6}(n-1)^7\nonumber
\end{eqnarray}
yield successive level crossings between regular $n$- and $(n+1)$-Rydberg atom chains at
\begin{equation}\label{cridet}
\Delta^n_{n+1}=\frac{C}{s_{||}^6}\left(n^7-(n-1)^7\right)\stackrel{n\gg1}{\rightarrow}7\frac{C}{s_{||}^6}n^6.
\end{equation}
Tuning $\Delta$ from negative to increasingly positive values, thus, induces stepwise changes in the many-body groundstate character from the initial state $|0\rangle$ to regular chains of $n$ excited Rydberg atoms ($|n\rangle$). Equation (\ref{cridet}) yields a well-defined large-$N$ limit, where the detuning 
\begin{equation}
\Delta_{\nu}=7C\rho^6\nu^6=7\frac{C}{a_{n+1}^6}
\end{equation}
determines the fraction $\nu=n/N$ of excited atoms and the Rydberg atom lattice spacing $a_n=s_{||}/(n-1)$. 

\begin{figure}[b!]
\begin{center}
\resizebox{0.99\columnwidth}{!}{\includegraphics{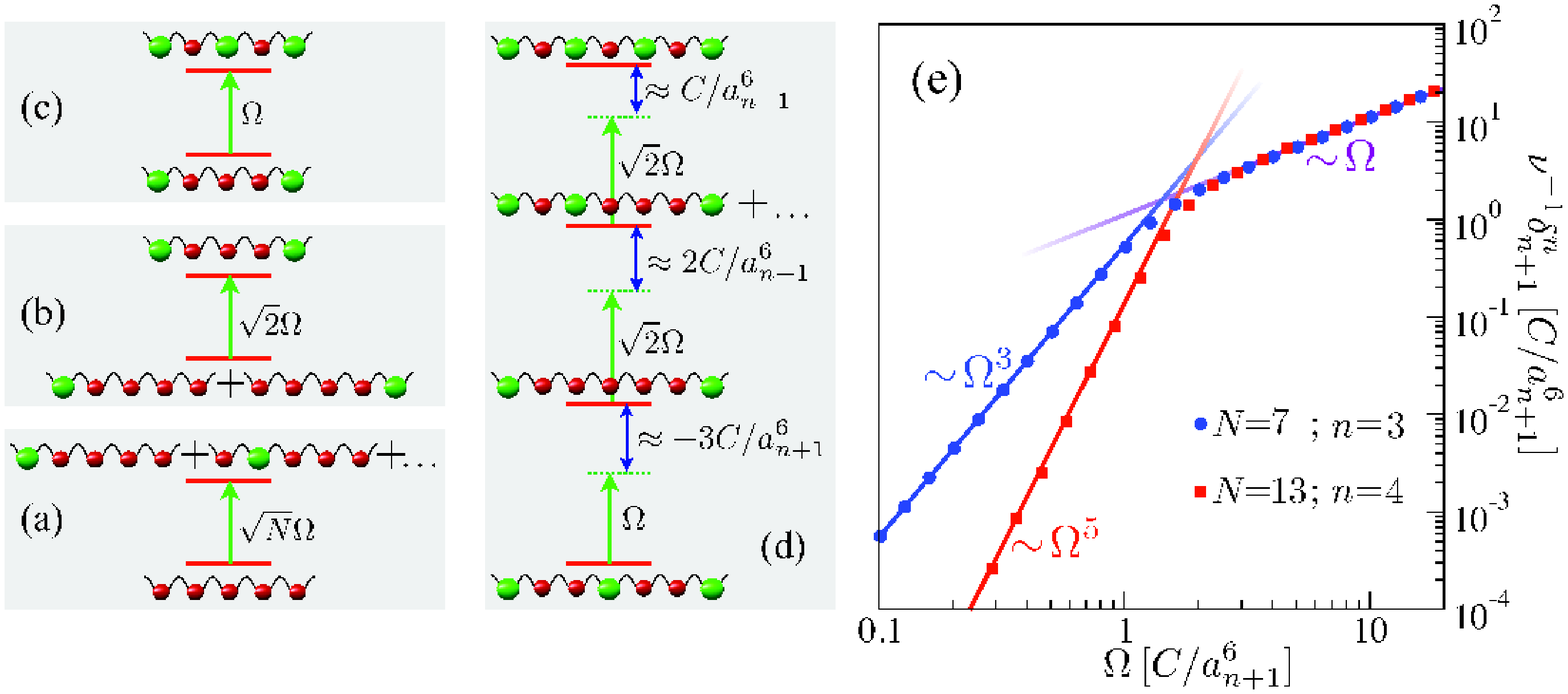}}
\caption{(a)-(c) Schematics of single-photon coupling between adjacent Rydberg lattice states $|n\rangle$ and $|n+1\rangle$ for $n<3$. (d) Schematics of the three-photon coupling for $n=3$. The intermediate states are off-resonant by $\sim C/a_{n+1}^6$. (e) Resulting energy gap for different $N$ and $n$ as a function of $\Omega$. The fitted power-laws demonstrate the transition from multi-photon to to sequential one-photon coupling, discussed in the text.\label{fig2}}
\end{center}
\end{figure}

Transitions between the crystal states are induced by the excitation laser, which couples adjacent $n$-manifolds, such that the degeneracies at $\Delta^n_{n+1}$ are lifted and replaced by avoided crossings of separation $\delta^n_{n+1}$. Adiabatic preparation of a given crystal state is possible provided that the pulse-evolution is slow compared to $1/\delta^n_{n+1}$.

At the first three crossings adjacent lattice states are directly coupled by the laser field (fig.\ref{fig2}a-c), such that the energy gap $\delta^n_{n+1}\sim\Omega$. For $n\ge 3$, however, there is no direct laser-coupling between adjacent levels (fig.~\ref{fig2}d). Instead, the gap arises from ($2n-3$)-photon transitions via off-resonant intermediate states detuned by $\sim C/a_{n+1}^6$. 

If at the time $t$ of a given $n\rightarrow(n+1)$ transition, $\omega_n=\Omega(t)/(Ca_{n+1}^{-6})\ll1$, the intermediate states, that couple $|n\rangle$ and $|n+1\rangle$ are far off-resonant and act as virtual levels for resonant multi-photon transitions. For a given Rydberg atom number $n$ these transitions involve $2n-2$ intermediate states such that the gap scales as
$\delta^n_{n+1}\sim\omega_n^{2n-3}\left({n}/({N-1})\right)^6{C}/{a^6}$
This exponential drop with increasing $n$, prevents the preparation of large crystals.  However, at larger Rabi frequencies, $\omega_n>1$, power broadening exceeds the intermediate state detunings. In this regime, consecutive crystal states are coupled by sequential one-photon transitions, which results in a linear $\Omega$-scaling of $\delta^n_{n+1}$.

This is confirmed by our numerical calculations, demonstrating the transition from multi- to single-photon coupling at $\Omega\sim C/a_n^6$ (see fig.~\ref{fig2}e). Beyond this critical Rabi frequency the gap scales as $\delta^n_{n+1}\sim\nu\Omega$,  ensuring a finite gap in the large-$n$ limit. Since the lattice spacing $a_n\approx1/(\nu\rho)$ is also independent of $s_{||}$, arbitrarily large Rydberg crystals should indeed be producible with finite peak Rabi frequencies $\Omega>C/a_n^6$.

We now focus on the excitation dynamics in frozen gases, of randomly spaced atoms.
To be specific, we consider a Bose-Einstein condensate, whose initial state $\prod_i\phi({\bf r}_i)\cdot\bigotimes_{i}|g_i\rangle$ is described by an axially symmetric Thomas-Fermi profile $\phi({\bf r})\propto \sqrt{1/4-(x^2+y^2)/s_{\perp}^2-z^2/s^2_{||}}$. Observables  are now determined by quantum average over the $N$-body spatial wave function and evaluated through Monte-Carlo integration. The fully correlated $N$-body wave function is obtained via a Hilbert space truncation approach that exploits the excitation blockade of closely spaced atoms \cite{yrp09}.

\begin{figure}[t!]
\begin{center}
\resizebox{0.7\columnwidth}{!}{\includegraphics{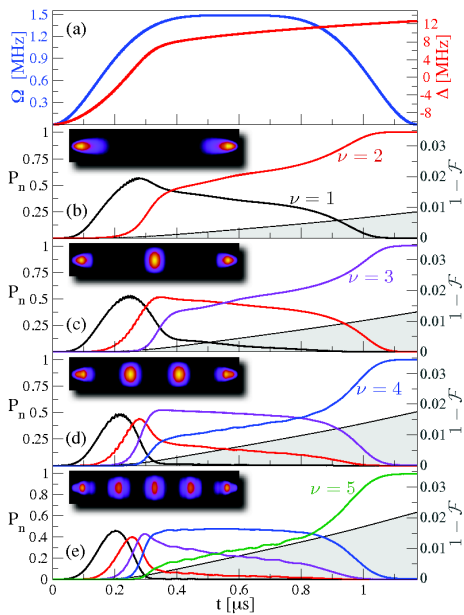}}
\caption{(a) Pulse envelope and chirp of the applied laser field, that excites a cigar-shaped Rb condensate to $65$s states. (b)-(e) Resulting time dependence of the probabilities $P_n$ to excite $n$ atoms in a gas of $300$ atoms and with a length of $s_{||}=15\mu$m (b), $22.5\mu$m (c), $35\mu$m (d), $45\mu$ (e). The insets show the resulting final Rydberg atom densities.\label{fig3}}
\end{center}
\end{figure}

The simple chain model, developed in the previous sections, provides a qualitative picture for the excitation dynamics in disordered systems (see fig.~\ref{fig1}d-e).
An important difference is that a disordered gas accommodates Rydberg atom configurations with continuously distributed interaction energies, leading to unavoidable non-adiabatic transitions within a given $n$-manifold at the end of the laser pulse. The resulting crystals are thus composed of localized collective excitations. 

\begin{figure}[b!]
\begin{center}
\resizebox{0.85\columnwidth}{!}{\includegraphics{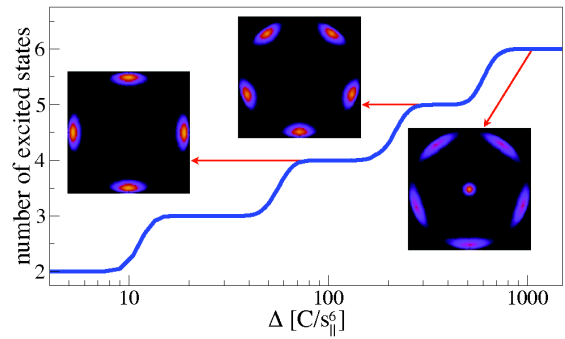}}
\caption{Dipole blockade staircase in a pancake-shaped condensate. The insets show Rydberg atom densities of the many-body groundstate at the indicated plateau values of the laser detuning $\Delta$.\label{fig4}}
\end{center}
\end{figure}

To demonstrate the experimental feasibility of the proposed scheme, we have simulated the dynamics of a cigar-shaped Rubidium condensate excited to $65s$ Rydberg states \cite{rlk07} with realistic laser pulses \cite{hrb07,ujh08} (fig.~\ref{fig3}a). For $s_{||}=15\mu$m, two excitations, localized at the edges of the cloud, are excited. Increasing $s_{||}$ gives smaller critical detunings $\Delta^n_{n+1}$ [see equation (\ref{cridet})], such that consecutively larger lattices are produced (fig.~\ref{fig3}b-\ref{fig3}e). Further increasing the cloud length $s_{||}$, we find a linear increase of the correlation length up to our largest systems size of $70\mu$m.
The depicted time evolution of the Rydberg atom number populations $P_n$  demonstrates that $-$ despite the disordered atom positions $-$ the gas is transfered to a state with precisely $n$ excitations, arranged on perfectly filled Rydberg crystals.

The described mechanism equivalently applies to higher dimensions and different geometries. Exemplary we show in fig.~\ref{fig4} the excitation-number staircase of a pancake-shaped condensate ($s_{\perp}\gg s_{||}$) along with some Rydberg atom densities of the many-body ground state. The depicted densities are obtained via a Monte Carlo average over randomly sampled atom positions, from which we obtain the lowest energy excited state configuration for a given $n$. To visualize the angular correlations, each individual configuration is rotated such that one excitation is centered at the bottom of the image and the  horizontal component of the Rydberg atoms' center of mass vanishes. Analogously to the one-dimensional case, Rydberg atoms are excited on regular lattice structures, which undergo a structural transition to a single-atom centered hexagon, with increasing $\Delta$ or  $s_{\perp}$. Increasing the system size $s_{\perp}$, thus, yields radial growth of large triangular crystals, with the lattice constant controllable by the detuning $\Delta$.

\begin{figure}[t!]
\begin{center}
\resizebox{0.99 \columnwidth}{!}{\includegraphics{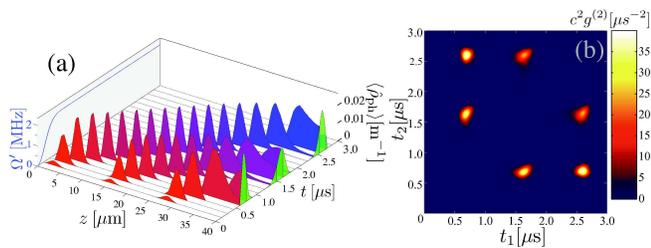}}
\caption{Light mapping of a three-Rydberg atom crystal with $s_{||}=30\mu$m and $s_{\perp}=4\mu$m, embedded in a condensate with a longitudinal and transverse size of $50\mu$m and $10\mu$m, respectively. (a) Pulse propagation dynamics, as described by the expectation value of the photon density operator $\hat{\rho}_{\rm ph}\propto {\hat{\bf E}}^{\dagger}{\hat{\bf E}}$. The outer-left blue curve shows the time evolution of the control Rabi frequency $\Omega^{\prime}$, while the right green area gives the resulting photon signal, detectable outside the atom cloud. (b) Corresponding two-photon correlation function $g^{(2)}=\langle\hat{\rho}_{\rm ph}(t_1)\hat{\rho}_{\rm ph}(t_2)\rangle$. Multiplication by the squared speed of light, $c$, yields a temporal density and ensures the correct  normalization $c^2\int g^{(2)} dt_1dt_2=n(n-1).$ \label{fig5}}
\end{center}
\end{figure}

Single-atom decay processes, such as black body radiation induced transitions and spontaneous decay typically occur on very long timescales of several $100\mu$ \cite{decay}. In the limit of small final decay probabilities, we can estimate its importance by following the coherent dynamics and determining the decay probability from the time-dependent number of excited atoms. As shown in Fig.\ref{fig3} the non-fidelity due to such processes is as small $\sim 0.02$, even for the largest considered Rydberg chain. As this values increases linearly with $n$, the preparation of rather large chains of up to $25$ atoms has a decay-limited fidelity of ${\mathcal F}\approx 0.9$. Cryogenic environments \cite{cgp05} reduce the value of $1-{\mathcal F}$ by a factor of $\sim3$ for the present parameters \cite{decay}, thus, permit high fidelity preparation of considerably larger crystals.
A less controllable decoherence source arises from atomic motion induced by the Rydberg-Rydberg atom interaction itself. We estimate its effect by the classical displacement $\Delta r\approx C r^{-7} M^{-1} \tau^2$ of an atom pair, initially separated by $r$, where $M$ is the mass of the atoms. For the parameters of fig.~\ref{fig3} ($r>10\mu$m) and $\tau=4.0\mu$s this gives a change in atom distances by less than a fraction of $10^{-3}$, leaving the excitation dynamics practically unaffected.

A number of techniques can be used to experimentally probe the Rydberg crystals. Absorption imaging of the groundstate atoms in optical lattices \cite{nlw07} could allow to detect excited atoms as defect spots at the corresponding lattice sites. Electric field ionization of Rydberg atoms combined with position sensitive MCP detectors and ion optics magnification \cite{rkt09} could allow to directly image the predicted $\mu$m-scale structures. Already measuring the final number distribution of Rydberg atoms \cite{crb05} would indirectly verify the production of the discussed crystals states, since they stand out due to a single number $n$ of excited atoms. 

Alternatively, the produced crystalline Rydberg atom configurations can be mapped onto correlated states of light and later probed via photon correlation measurements.
Here, a strong $\pi$-pulse is used to coherently transfer the Rydberg lattice to a crystal state of atoms in another hyperfine groundstate $|g^{\prime}\rangle$.
A second laser with Rabi frequency $\Omega^{\prime}(t)$ drives the transition between $|g^{\prime}\rangle$ and a low-lying excited state $|e^{\prime}\rangle$, which is coupled to $|g\rangle$ by a quantum field $\hat{\bf{E}}({\bf r},t)$. 
Analogous to light storage schemes in an EIT medium \cite{fl00} the initially prepared quantum state of atomic excitations can be transfered to the radiation field $\hat{\bf E}$ by adiabatically turning on the control field $\Omega^{\prime}$. 
Fig.~\ref{fig5} shows the resulting transfer dynamics for the specific example of a three-atom chain, embedded in a larger condensate \footnote{Such embedded crystal states reduce effects of density inhomogeneities and can be produced via crossed-beam, two-photon excitation \cite{rla08}.}.
The collective nature of the Rydberg excitations not only provides enhanced coupling to light but also leads to highly directed photon emission along the total wave vector of the applied laser fields ${\bf k}_{\rm tot}$ (fig.~\ref{fig5}a). The Rydberg chain is mapped onto a clear triple-peak structure, with well resolvable delay times of $\sim1\mu$s. The corresponding second order correlations, shown in fig.~\ref{fig5}b, clearly reveal the single-photon nature of each pulse, as characterized by the absence of equal-time photon signals. Since there is a one-to-one mapping between the photon density $\hat{\rho}_{\rm ph}$ and the initial Rydberg state density operator \cite{fl00}, the depicted time correlations directly reflect the strongly correlated nature of the prepared many-body state. In this way, correlation measurements, provide a direct probe for (im)perfection of the underlying Rydberg lattice. 

In summary, we have shown that the excited-state number of laser-driven Rydberg gases exhibits a stepwise increase as a function of the laser frequency, which bears analogies to the Coulomb blockade staircase known from nano-scale solid state devices. Our dynamical calculations show that frequency chirps allow to adiabatically ascend this dipole blockade staircase, thereby forming ordered structures of Rydberg atoms out of otherwise  disordered ensembles. 
Such crystals naturally realize a network of localized collective excitations without the need of extended micro-trap arrays.
While our simulations are limited to comparably small excitation numbers, the obtained scaling laws suggest that large crystals can be prepared under typical experimental conditions. Chirped excitation of lattice-confined atoms to large Rydberg atom fractions would provide a promising experimental scenario to study magnetism of long-range interacting spins and explore their quench dynamics through quantum phase transitions. We described how quantum optical imaging technique could be used as a real-space probe of spatial correlations between the effective spins, thereby providing a novel source of non-classical light.

This work was supported by NSF, DARPA, AFOSR MURI and Packard Foundation.

\end{document}